\documentclass[%
 superscriptaddress,
preprint,
 amsmath,amssymb,
prx,
floatfix,
]{revtex4-2}

\newcommand{\Er}{$^{167}$Er$^{3+}$}
\newcommand{\Ers}{$^{167}$Er$^{3+}$ }
\newcommand{\YSO}{Y${_2}$SiO${_5}$ }
\newcommand{\YSOns}{Y${_2}$SiO${_5}$}
\newcommand{\aSi}{$\alpha$Si}
\newcommand{\aSis}{$\alpha$Si }

\usepackage{graphicx}% Include figure files
\usepackage{dcolumn}% Align table columns on decimal point
\usepackage{bm}% bold math
\begin{document}

\title{Multifunctional on-chip storage at telecommunication wavelength for quantum networks}

\author{Ioana Craiciu}
	\thanks{These two authors contributed equally to this work.}
	\affiliation{Kavli Nanoscience Institute and Thomas J. Watson, Sr., Laboratory of Applied Physics, California Institute of Technology, Pasadena, California 91125, USA}
	\affiliation{Institute for Quantum Information and Matter, California Institute of Technology, Pasadena, California 91125, USA}
\author{Mi Lei}
	\thanks{These two authors contributed equally to this work.}
	\affiliation{Kavli Nanoscience Institute and Thomas J. Watson, Sr., Laboratory of Applied Physics, California Institute of Technology, Pasadena, California 91125, USA}
	\affiliation{Institute for Quantum Information and Matter, California Institute of Technology, Pasadena, California 91125, USA}
\author{Jake Rochman}
	\affiliation{Kavli Nanoscience Institute and Thomas J. Watson, Sr., Laboratory of Applied Physics, California Institute of Technology, Pasadena, California 91125, USA}
	\affiliation{Institute for Quantum Information and Matter, California Institute of Technology, Pasadena, California 91125, USA}
\author{$\mbox{John G. Bartholomew}$} \altaffiliation[Currently at: ]{School of Physics, The University of Sydney, Sydney, New South Wales 2006, Australia\\
The University of Sydney Nano Institute, The University of Sydney, NSW 2006, Australia}
	\affiliation{Kavli Nanoscience Institute and Thomas J. Watson, Sr., Laboratory of Applied Physics, California Institute of Technology, Pasadena, California 91125, USA}
	\affiliation{Institute for Quantum Information and Matter, California Institute of Technology, Pasadena, California 91125, USA}
\author{Andrei Faraon}
 	\email[Corresponding author: ]{faraon@caltech.edu}
	\affiliation{Kavli Nanoscience Institute and Thomas J. Watson, Sr., Laboratory of Applied Physics, California Institute of Technology, Pasadena, California 91125, USA}
	\affiliation{Institute for Quantum Information and Matter, California Institute of Technology, Pasadena, California 91125, USA}

\date{\today}

\begin{abstract}

Quantum networks will enable a variety of applications, from secure communication and precision measurements to distributed quantum computing. Storing photonic qubits and controlling their frequency, bandwidth and retrieval time are important functionalities in future optical quantum networks. Here we demonstrate these functions using an ensemble of erbium ions in yttrium orthosilicate coupled to a silicon photonic resonator and controlled via on-chip electrodes. Light in the telecommunication C-band is stored, manipulated and retrieved using a dynamic atomic frequency comb protocol controlled by linear DC Stark shifts of the ion ensemble's transition frequencies. We demonstrate memory time control in a digital fashion in increments of 50 ns, frequency shifting by more than a pulse-width ($\pm39$ MHz), and a bandwidth increase by a factor of three, from 6 MHz to 18 MHz. Using on-chip electrodes, electric fields as high as 3 kV/cm were achieved with a low applied bias of 5 V, making this an appealing platform for rare earth ions, which experience Stark shifts of the order of 10 kHz/(V/cm).
\end{abstract}

\maketitle

\section{Introduction}

Optical quantum memories will enable long distance quantum communication using quantum repeater protocols \cite{Briegel1998,Kimble2008,Heshami2016}. A quantum memory device which can control the bandwidth and frequency of stored light is additionally useful, as it can interface between optical elements which have different optimal operating points. Erbium-doped materials are a promising solid-state platform for ensemble-based optical quantum memories because of their long-lived optical transition in the telecommunication C-band that is highly coherent at cryogenic temperatures \cite{Bottger2006,Bottger2009}. This allows for integration of memory systems with low-loss optical fibers, opening up opportunities for repeaters over continental distances, as well as integration with silicon photonics \cite{Miyazono2017,Dibos2018} one of the most advanced platforms for integrated photonics. Spin transitions in \Er-doped yttrium orthosilicate (\Er:\YSOns) have also been shown to have long relaxation and coherence lifetimes at cryogenic temperatures and high magnetic fields \cite{Rancic2018} which opens the possibility for long term spin wave memories. 

There have been several demonstrations of optical storage in erbium-doped materials \cite{Saglamyurek2015,Askarani2019,Lauritzen2010,Lauritzen2011,Dajczgewand2014,Craiciu2019}, including storage at the quantum level \cite{Saglamyurek2015,Askarani2019,Craiciu2019}, and on-chip storage \cite{Askarani2019,Craiciu2019}. These results are part of a larger body of optical quantum memory research \cite{Brennen2015,Heshami2016}, using rare-earth-ion-doped crystals \cite{Hedges2010,Schraft2016,Holzapfel2020,Heinze2013}, atomic gases \cite{Hsiao2018,Wang2019}, and single atoms or defects \cite{Specht2011,Bhaskar2020}. In parallel with efforts to increase the efficiency \cite{Hsiao2018} and storage time \cite{Heinze2013,Holzapfel2020} of quantum memories, several works have focused on new types of multifunctional devices \cite{Hosseini2009,Fisher2016,Mazelanik2019} in which control fields are used to modify the state of the light during storage. 

In many quantum repeater protocols \cite{Wu2020}, quantum memories act as interfaces between emitters such as quantum dots \cite{Wang2019dot} or individual atoms \cite{Dibos2018}. Dynamic control of the optical pulses stored in these memories can correct for differences between individual emitters, leading to higher indistinguishably for Bell State measurements at the entanglement swapping stage of quantum repeater protocols \cite{Briegel1998}. In addition, with control over the frequency of stored light, one can map an input mode to a different output mode in a frequency multiplexed quantum memory, which enables quantum networks with fixed-time quantum memories \cite{Sinclair2014}.

In this work, we use a silicon resonator evanescently coupled to \Er:\YSO ions and gold electrodes to realize a multifunctional on-chip device which can not only store light, but also dynamically modify its frequency and bandwidth. Electrodes create a DC electric field that can be rapidly switched, which enables control of the \Ers ions' optical transition frequency via the DC Stark shift \cite{Macfarlane2007}. Using a resonator increases the interaction between light and the ion ensemble, allowing on-chip implementation of the atomic frequency comb (AFC) memory protocol \cite{Afzelius2010}. This protocol allows multiplexing in frequency, which offers a significant advantage in quantum repeater networks \cite{Simon2007}. Additionally, the on-chip electrodes are patterned close together to achieve the high electric fields required for Stark shift control with CMOS compatible, applied voltages. We demonstrate dynamic control of memory time in a digital fashion, as well as modification of the frequency and bandwidth of stored light. 

\section{Hybrid \lowercase{$\alpha$}S\lowercase{i}-$^{167}$E\lowercase{r}$^{3+}$:Y$_2$S\lowercase{i}O$_5$ resonator with electrodes}

\begin{figure*}
\includegraphics[width=6.5in]{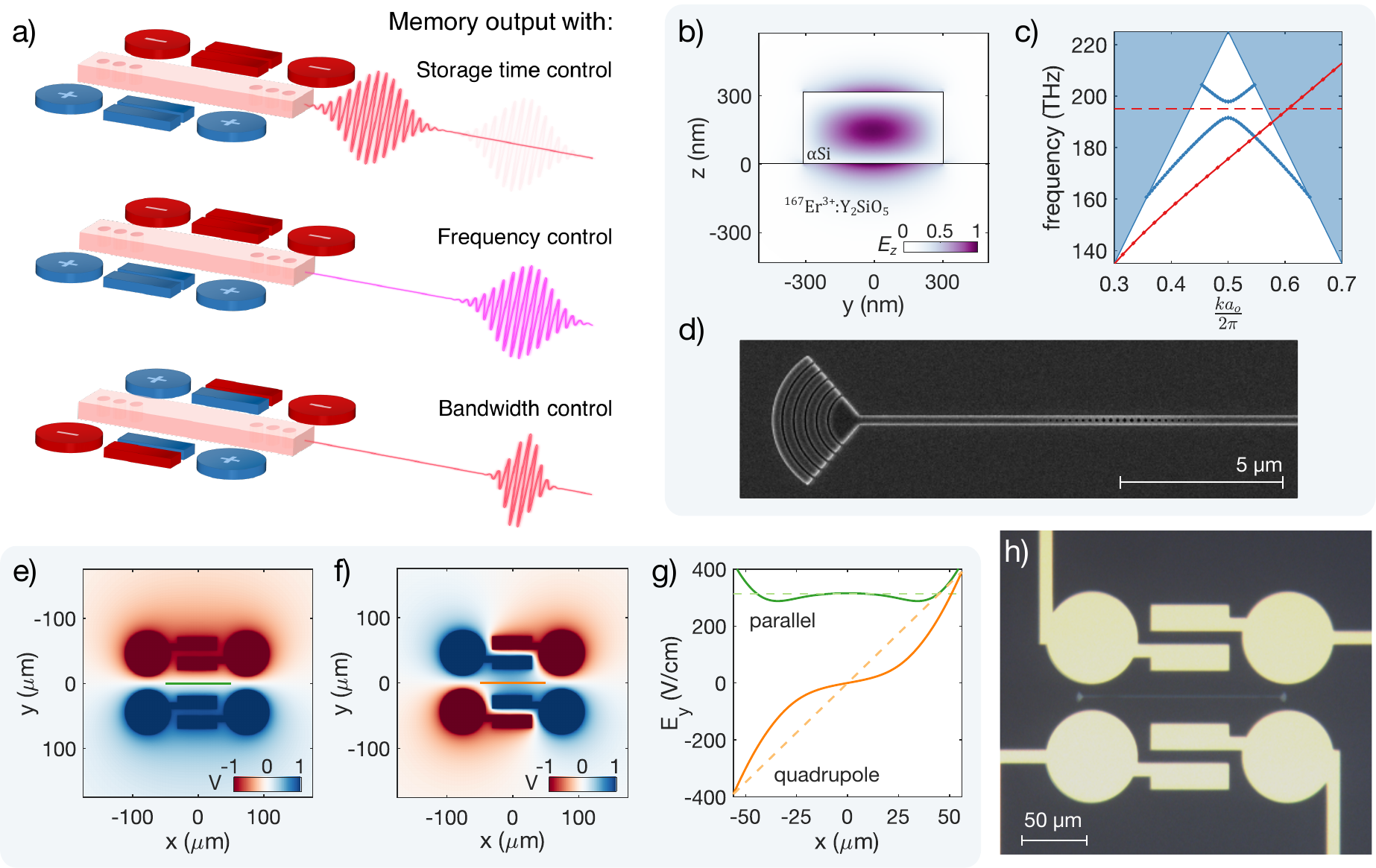}
\caption{Multifunctional quantum storage device. a) Schematic of device functionality showing the optical resonator (pink), electrodes (blue and red) and memory output. b-d) Hybrid optical resonator comprised of an amorphous silicon (\aSi) waveguide on \Er:\YSO with photonic crystal mirrors on either end. b) cross-section of waveguide (black outline) showing 2D finite element simulation of the transverse-magnetic waveguide mode. Purple-white gradient indicates the $E_z$ component of the optical field. c) Band diagram showing waveguide mode (solid red line), band gap of a photonic crystal mirror (solid blue lines), and the design frequency of 195 THz (dashed red line). Blue areas indicate the \YSO light cone containing extended modes propagating in bulk \YSO (both sides for photonic crystal, left side only for waveguide). d) Scanning electron micrograph showing a grating coupler and photonic crystal mirror including tapered sections on either side to reduce scattering \cite{Deotare2009}. e-g) 3D finite element simulation of on-chip electrodes. e-f) 2D slice at $z=0$ showing electric potential (blue-red gradient) in the parallel (e) and quadrupole (f) biasing configurations. g) Electric field $E_y(x)$ along optical resonator in the parallel (green solid line) and quadrupole (orange solid line) configurations; $E_y(x)$ was measured at $z=0$, $y=0$, $-56\:\mu\mathrm{m}<x<56\:\mu\mathrm{m}$ (green and orange lines in (e) and (f)); dashed lines indicate ideal parallel (green) and quadrupole (orange) electric field distributions. h) Optical micrograph showing an optical resonator, gold electrodes, and gold wires for electrical contact.}\label{fig:device}
\end{figure*}

The multifunctional device consists of an optical resonator coupled to \Er:\YSO ions between gold electrodes. Using the AFC quantum storage protocol \cite{Afzelius2009} and the ions' Stark shift, light can be stored and manipulated in this device. Figure \ref{fig:device}a shows a schematic of the device and the three functionalities demonstrated in this work: memory time control, frequency control, and bandwidth control. Different electric field configurations are created by applying a positive (blue) or negative (red) bias to each electrode. For the true device dimensions, see the micrograph in Fig$.$ \ref{fig:device}h.

The optical resonator used in this work is a Fabry-Perot resonator comprised of a 100 $\mu$m amorphous silicon (\aSi) waveguide on \Er:\YSO with photonic crystal mirrors on either end. Figures \ref{fig:device}b-d show simulations and micrographs of this resonator. The waveguide is $h=310$ nm tall and $w=605$ nm wide. Ten percent of the energy of the transverse-magnetic optical waveguide mode penetrates into the \Er:\YSO and evanescently couples to the \Ers ions. Photonic crystal mirrors on either side are formed by a repeating pattern of elliptical air holes in the \aSis waveguide with period $a_\circ=370$ nm. A grating coupler is used to couple light from a free-space mode into and out of the resonator. The amorphous silicon resonator is fabricated on top of an \Er:\YSO chip using a deposition and etching process similar to Ref$.$ \cite{Miyazono2017}. The \Er:\YSO substrate is doped with isotopically purified \Ers ions at 135 ppm, measured by secondary ion mass spectrometry, and cut perpendicular to the $D_1$ crystal axis, such that the electric field of the transversal magnetic (TM) optical mode is polarized along this axis. 

Quality factors of up to $10^5$ were measured for weakly coupled resonators, where the photonic crystal mirrors on both sides were designed to be highly reflective. The device used in this work is made one-sided for more efficient quantum storage \cite{Afzelius2009,Craiciu2019} by using fewer photonic crystal periods in one mirror to make it less reflective. Light is sent into and measured from the side with the lower reflectivity mirror. The intrinsic quality factor for this device is also lower than the weakly coupled resonators, leading to a quality factor of $3\times10^4$ and a coupling ratio of $\kappa_\mathrm{in}/\kappa=0.2$, where $\kappa_\mathrm{in}$ is the coupling rate through the lower reflectivity mirror and $\kappa$ is the total decay rate \cite{Groblacher2013}.

Electrodes are used to apply electric fields to those ions coupled to the optical resonator. There are four independently biased gold electrodes, each comprised of a 70 $\mu$m diameter circle connected to a $20\:\mu\mathrm{m} \times 60\:\mu\mathrm{m}$ rectangle. They are patterned onto the \Er:\YSO after the \aSis resonators using electron-beam lithography followed by electron-beam gold evaporation and lift-off. Figures \ref{fig:device}e-g show simulations of the two electrode biasing configurations: parallel, which applies a nearly constant electric field to all ions ($E(x)=a$), and quadrupole, which applies an electric field gradient along the resonator (approximating $\frac{\mathrm{d}E(x)}{\mathrm{d}x}=b$), where $a$ and $b$ are constants. The electrode geometry was designed to best approximate these two electric field profiles with four independently biased electrodes, while providing a large electric field for a given applied bias ($E/V$). In the \Er:\YSO region where ions are coupled to the optical mode, the $E_y$ component of the electric field is dominant ($E_y \gg E_x,E_z$), and it does not vary significantly in the $z$ and $y$ directions. Therefore only $E_y(x)$, which is aligned to the $b$-axis of the \YSO crystal, is considered.

The device is thermally connected to the coldest plate of a dilution refrigerator, the temperature of which is $\sim 100$ mK. A static magnetic field of 0.98 T is applied along the \YSO $D_1$-axis with a superconducting electromagnet. Trim coils are used to cancel any magnetic field component along the $b$-axis. The remainder of the measurement setup was similar to the one described in Ref$.$ \cite{Craiciu2019}.

\section{DC Stark shift in $^{167}$E\lowercase{r}$^{3+}$:Y$_2$S\lowercase{i}O$_5$}

Er$^{3+}$:\YSO has been extensively studied for quantum applications \cite{Bottger2006,Bottger2009,Hastings-Simon2008,Lauritzen2010,Lauritzen2011,Dajczgewand2014,Dibos2018,Craiciu2019,Rakonjac2020}, including demonstrations of AFC storage \cite{Lauritzen2010,Lauritzen2011,Craiciu2019}. Erbium ions substitute for yttrium ions in \YSO in 2 crystallographic sites, each of which has four different orientations due to the $C^6_{2h}$ crystal symmetry \cite{Sun2008}. In this work, we use crystallographic site 2, which has an optical transition near $1539$ nm \cite{Bottger2006}. \Ers has a nuclear spin $I=7/2$, which together with an effective electron spin, leads to 16 hyperfine levels in both the optical ground and excited states. At high fields and low temperatures, the lowest 8 ground-state hyperfine levels are long-lived \cite{Rancic2018,Craiciu2019}, enabling the spectral holeburning that is required to create atomic frequency combs. 

Dynamic control is enabled by the DC Stark shift. When a rare earth ion in a crystal interacts with a DC electric field $\vec{E}$, its optical transition frequency is shifted due to the difference between the permanent electric dipole moments in the optical excited and optical ground states $\delta\vec{\mu}=\vec{\mu}_\mathrm{e}-\vec{\mu}_\mathrm{g}$. For non-centrosymmetric sites such as the yttrium sites in \YSO for which Er$^{3+}$ ions substitute, the linear Stark shift term $\delta f = - \frac{1}{h}\, \delta\vec{\mu} \cdot \hat{L} \cdot \vec{E}$ dominates, where $\hat{L}$ is the local field correction tensor \cite{Macfarlane2007}. 

The Stark shift is dependent on the orientation of the applied field relative to $\delta\vec{\mu}$ \cite{Graf1997}. Without knowing $\delta \vec{\mu}$ or $\hat{L}$, the Stark shift can be empirically characterized for an electric field applied in a particular direction $\hat{n}$ using the Stark shift parameter $s_{\hat{n}}$ given $\delta f = s_{\hat{n}} E_{\hat{n}}$. We measured $s_{\hat{n}}=11.8 \pm 0.2$ kHz/(V/cm) for $\hat{n}$ nominally aligned with the \YSO crystal $b$-axis (see Appendix \ref{sec:Starkshift}).

In an ensemble of \Er:\YSO ions, four different Stark shifts will be observed for an arbitrary electric field due to the four orientations of each crystallographic site \cite{Sun2008}. For electric fields parallel or perpendicular to the $b$-axis, the Stark shifts of the four subclasses are pair-wise degenerate, resulting in two equal and opposite Stark shifts $\delta f_{\pm}=\pm s E$.  In this work, all electric fields are applied parallel to the $b$-axis, so we will simply refer to two \Ers subclasses.

\section{Atomic frequency comb storage with dynamic memory time control}

After a photon is absorbed by an ensemble of ions, the ensemble of ions is described by a Dicke state \cite{Afzelius2009}:

\begin{equation}
   \left|\Psi\right\rangle = \sum_{j=1}^{N_\mathrm{ions}} c_j\, e^{i2\pi \left(f_j+\,\delta f_j(t)\right)\,t}\,e^{-i k \vec{r}_j}\left|0 ... 1_j ... 0_N\right\rangle. \label{eq:Dicke}
\end{equation}

Each ion has a different transition frequency $f_j$ and position $\vec{r}_j$. For AFC storage, the transition frequencies $\{f_j\}$ form a frequency comb with period $\Delta$. When a photon is absorbed at $t=0$, the ensemble of ions first dephase then rephase every $t=\frac{m}{\Delta}$, $m \in \mathbb{N}$, leading to a coherent re-emission of the light \cite{Afzelius2009}. A Stark shift $\delta f_j(t)$ enables dynamic control of light stored in the AFC by changing the optical transition frequencies of the ions. $\delta f_j(t)$ can be varied over time by changing the amplitude of the applied electric field (slowly relative to optical frequencies). This enables two types of control: electric field pulses applied between the absorption and emission of light modify the phase of the output, while electric field pulses applied during emission of light modify the frequency profile of the output light.

To achieve dynamic control of storage time, the electrodes are biased in a parallel configuration as shown in the top panel of Figure \ref{fig:device}a. When an electric pulse is applied, the two \Er:\YSO subclasses experience equal and opposite frequency shifts, $\pm\delta f(t)=\pm s_b E$. By appropriately choosing the length in time $t$ and amplitude $E$ of the electric pulse, a $\pi$ phase difference between subclasses can be introduced $\pi=2\pi \times (+s_b E - (- s_b E))\times t$, which will prevent any coherent emission from the ensemble. An equal and opposite electric pulse can then rephase the two subclasses, and allows coherent re-emissions from the AFC. This procedure of dephasing and rephasing the ensemble works even if the electric field distribution is not perfectly homogeneous, as shown in the context of Stark Echo Modulation Memory in Reference \cite{Arcangeli2016}. Recently, dynamic control of memory time in AFC was demonstrated using this same procedure in Pr$^{3+}$:\YSO \cite{Horvath2020}. Reference \cite{Lauritzen2011} proposed a similar protocol but using an electric field gradient.

\begin{figure}
\centering\includegraphics[width=2.9in]{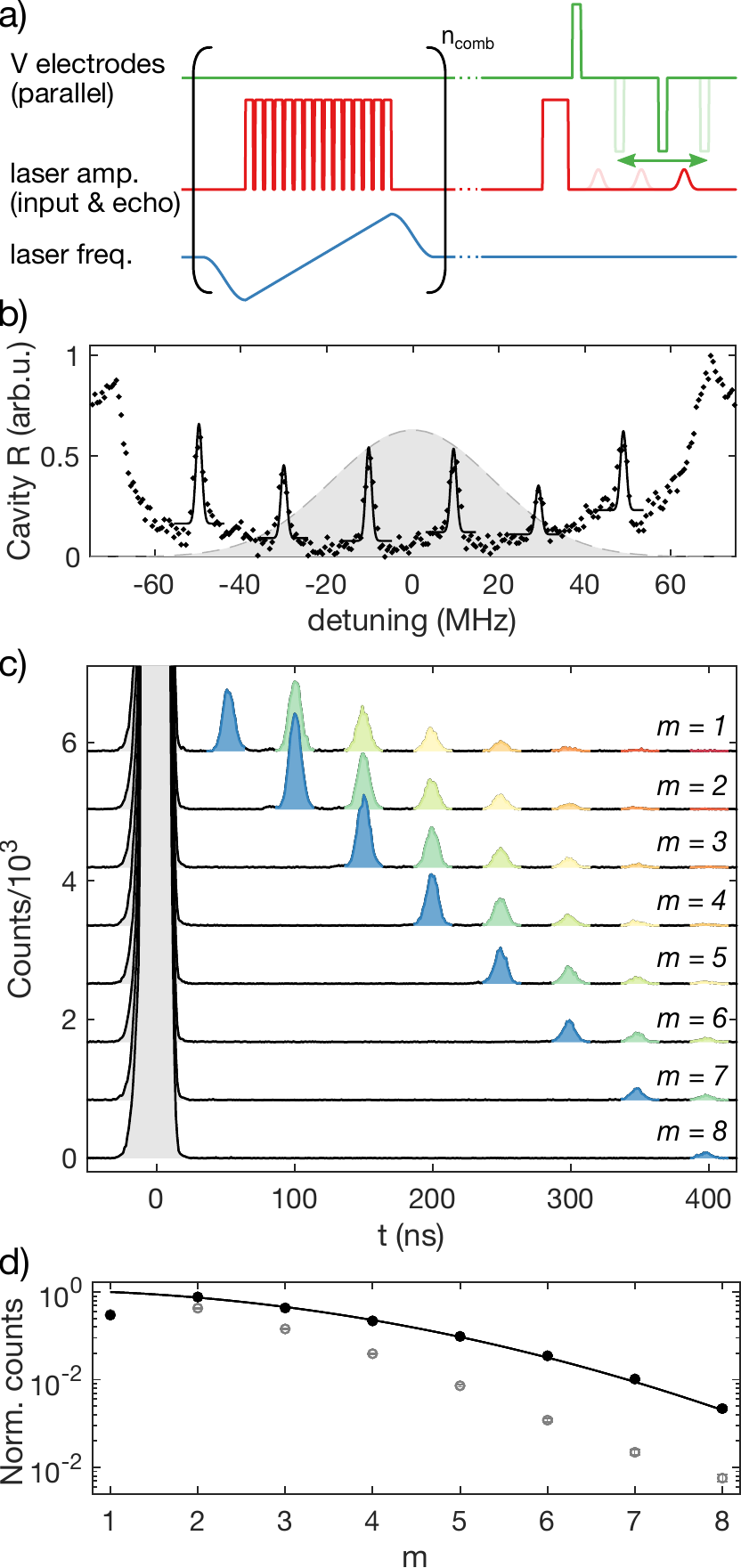}
\caption{(Color online) AFC storage with dynamic memory time control. a) Pulse sequence (not to scale, details in main text). b) Atomic frequency comb. Cavity reflectance (black points) and fit to six Gaussians (solid black lines). All teeth are fit together, with the finesse fixed to the value from (d). Detuning is measured from 194822 GHz. Grey Gaussian with dashed outline represents the input pulses in frequency space. c) Emission of stored light at different times $t_\mathrm{memory}=\frac{m}{\Delta}$. Partly reflected input pulse are shown in grey at $t=0$. On-demand memory outputs are shown in blue (darkest shade). Subsequent emissions (green to red) are discussed in the main text. Electric pulses are not shown. d) Energy emitted in the time bin at $t=\frac{m}{\Delta}$ for each value of $m$. Black data points represent the normalized counts when all previous emissions are suppressed (blue pulses in (c)). Grey data points represent normalized counts when previous emissions are not suppressed (all pulses on line $m=1$ in (c)). Error bars, representing $\sqrt{N_\mathrm{counts}}$, are smaller than the markers. Solid line is a fit to theory, fitting only for comb finesse.}\label{fig:killnr}
\end{figure}

The pulse sequence used to achieve dynamic control of AFC storage is shown in Figure \ref{fig:killnr}a. Not shown is the initialization to move most of the population into one hyperfine state, which is performed before every experiment \cite{Craiciu2019,Rancic2018}. First, an AFC with period $\Delta$ is created by repeatedly burning away population between the teeth of the comb, $n_\mathrm{comb} = 20$ times. Then, an input pulse indicated by the red laser pulse is sent into the resonator at $t=0$ and is absorbed by the AFC. Shown in light red are possible emissions corresponding to rephasing events of the AFC at times $t=\frac{m}{\Delta}$. Without electric field control, the output of the memory (the first and largest emission), would be centered at $t=\frac{1}{\Delta}$ ($m=1$). The schematic shows instead an emission in red at $t=\frac{3}{\Delta}$ ($m=3$), obtained
when a first electric pulse is applied before the first emission and a compensating pulse is applied immediately before the third emission.

Figure \ref{fig:killnr}b shows the AFC used in this experiment. The period of the comb, extracted from the fit, is $19.7 \pm 0.1$ MHz, which corresponds to a minimum storage time of $t=\frac{1}{\Delta}=50$ ns. Figure \ref{fig:killnr}c shows dynamically controlled storage for various values of $m$. Two electric pulses were used to control memory time. The first was a 10 ns long pulse with amplitude 2.0 kV/cm centered at $t_\mathrm{pulse \,1}=25$ ns. The second was 10 ns long with an opposite amplitude of -2.0 kV/cm, and its center position was varied as $t_\mathrm{pulse \,2} = 25$ ns $+\, (m-1) \times 50$ ns to allow the emission at $t_\mathrm{memory}=\frac{m}{\Delta}$. For the $m=1$ case, no electric pulses were applied. Between the two electric pulses, emission was suppressed down to the dark counts level, a factor of 100 lower than peak emission counts. The presence of multiple smaller pulses following the output pulse is a feature of the high finesse and low efficiency of the memory (see Appendix \ref{sec:efficiency_theory}). For higher efficiency, high finesse AFCs, subsequent emissions are significantly suppressed \cite{Horvath2020}.

Figure \ref{fig:killnr}d shows the energy emitted in the $m^{th}$ time bin for $t_\mathrm{memory}=\frac{m}{\Delta}$. The data is fit to the dephasing term in the theoretical storage efficiency for a comb with Gaussian teeth: $\mathrm{exp}\left(-\frac{\pi^2}{2 \mathrm{ln} 2}\frac{m^2}{F^2}\right)$ \cite{Afzelius2009,Lauritzen2011}, where $F=\Delta/\gamma$ is the comb finesse, and $\gamma$ is the full-width at half maximum (FWHM) of each tooth. The $m=1$ data point is excluded from the fit because the approximately 100 ns dead time of the single photon detector after the input pulse is thought to lead to undercounting in that time bin. A comb finesse of $F=12.2 \pm 0.2$ ($\gamma=1.6$ MHz) is extracted from this fit. This corresponds to a $1/e$ point of 240 ns ($m=4$ for digital storage time). To improve on this scaling requires a smaller tooth width $\gamma$. The grey data in Fig$.$ \ref{fig:killnr}d show the total counts in the $m^{th}$ time bin when the previous output pulses are not suppressed.  

\section{Dynamic frequency control}

\begin{figure}
\includegraphics[width=3in]{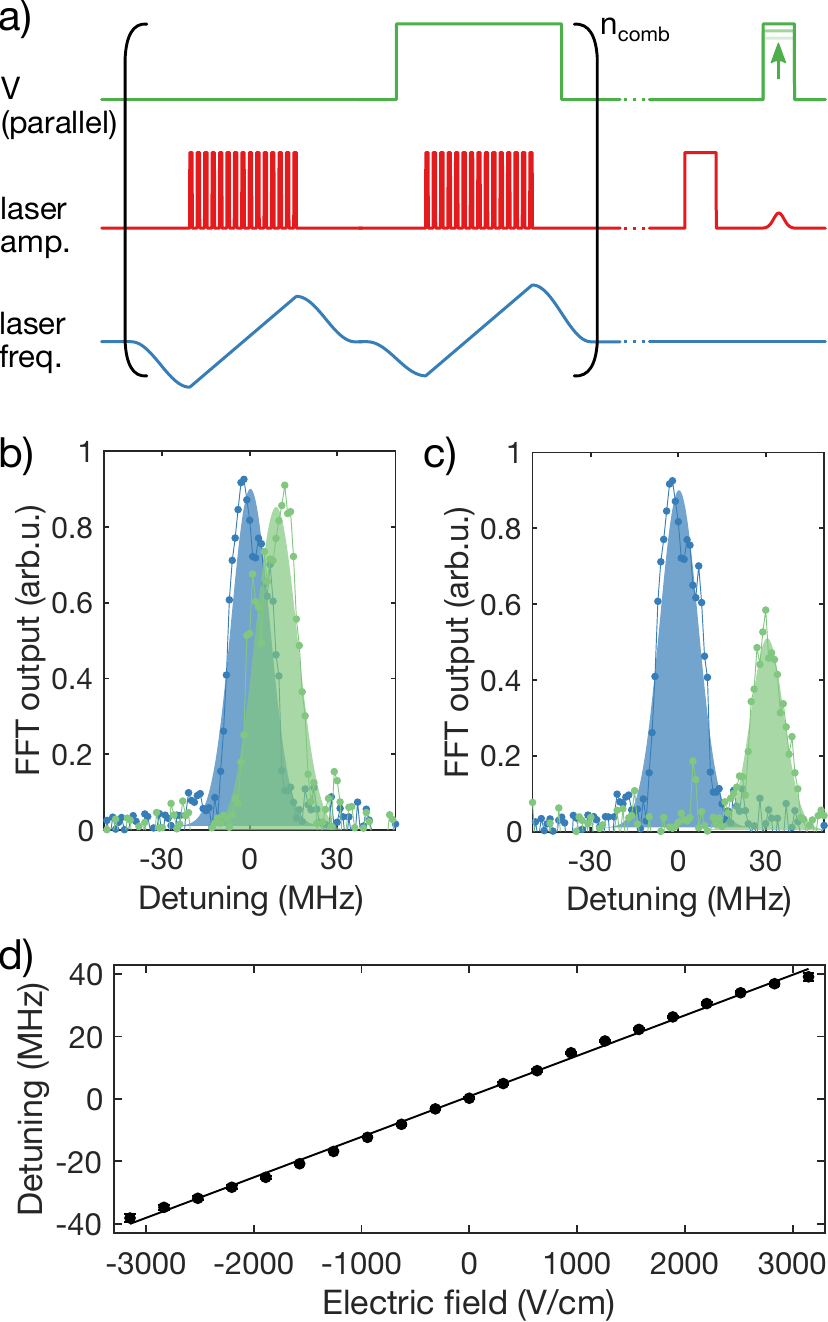}
\caption{(Color online) AFC storage with frequency control. a) Pulse sequence (not to scale, details in main text). b-c) Examples of AFC output pulses with (green, lighter) and without (blue, darker) a frequency shift. Filled in area is a Gaussian fit to the data (circles). Detuning is measured from 194822 GHz. Frequency shifts of 9 MHz (b) and 30 MHz (c) are shown. d) The output detuning as a function of electric field applied during emission. Circles are the centers of Gaussian fits (as shown in (b-c)). Error bars, which are smaller than the markers, are $95\%$ confidence intervals for those fits. The solid line is a linear fit to the data, yielding a slope of $13.0 \pm 0.3$ \mbox{kHz/(V/cm)}, similar to the Stark shift value measured by holeburning spectroscopy (Appendix \ref{sec:Starkshift})} \label{fig:freq}
\end{figure}

The frequency of light stored in an AFC can be dynamically modified during emission. The atomic frequency comb is shifted in frequency during the emission of stored light by biasing the electrodes in the parallel configuration as shown in the middle panel of Figure \ref{fig:device}a. The pulse sequence used to achieve AFC storage with frequency control is shown in Figure \ref{fig:freq}a. The first step is to eliminate one of the two \Ers subclasses from the spectral window, leaving only ions which experience a positive 
Stark shift, $\delta f_+ = +s_b E$ (the choice of subclass is arbitrary). This is accomplished using a two-part comb burning procedure. With the first burning step, a normal AFC containing both subclasses is created using a sequence of laser pulses. For the second burning step, the two subclasses are split by $\Delta/2$ using a parallel electric field, and a similar sequence of laser pulses is used, but with a frequency shift of $\Delta/4$. This burns away ions with a negative shift $\delta f_-=-s_b E$. 

Repeating the comb burning procedure $n_\mathrm{comb}=5$ times, an AFC with width 145 MHz, and a period $\Delta = 5$ MHz is created. An input pulse is sent in and the rephasing of the AFC causes an emission at $t=\frac{1}{\Delta} = 200$ ns. During this emission, an electric field pulse with amplitude $E_\mathrm{pulse}$ applied in the parallel configuration will cause the ions to emit with a frequency shift of $f = +s_b E_\mathrm{pulse}$. Figures \ref{fig:freq}b-c show the light emitted from the memory, with and without a frequency shift. A heterodyne measurement is used to measure the frequency of the output pulse directly. Figure \ref{fig:freq}d shows the linear frequency shift as a function of electric field. The decrease in output amplitude with frequency shift evident in Figures \ref{fig:freq}b-c is mainly due to the inhomogeneity of Stark shifts experienced by the ions (see Appendix \ref{sec:Starkshift}). 

\section{Dynamic bandwidth control}

\begin{figure}
\includegraphics[width=3in]{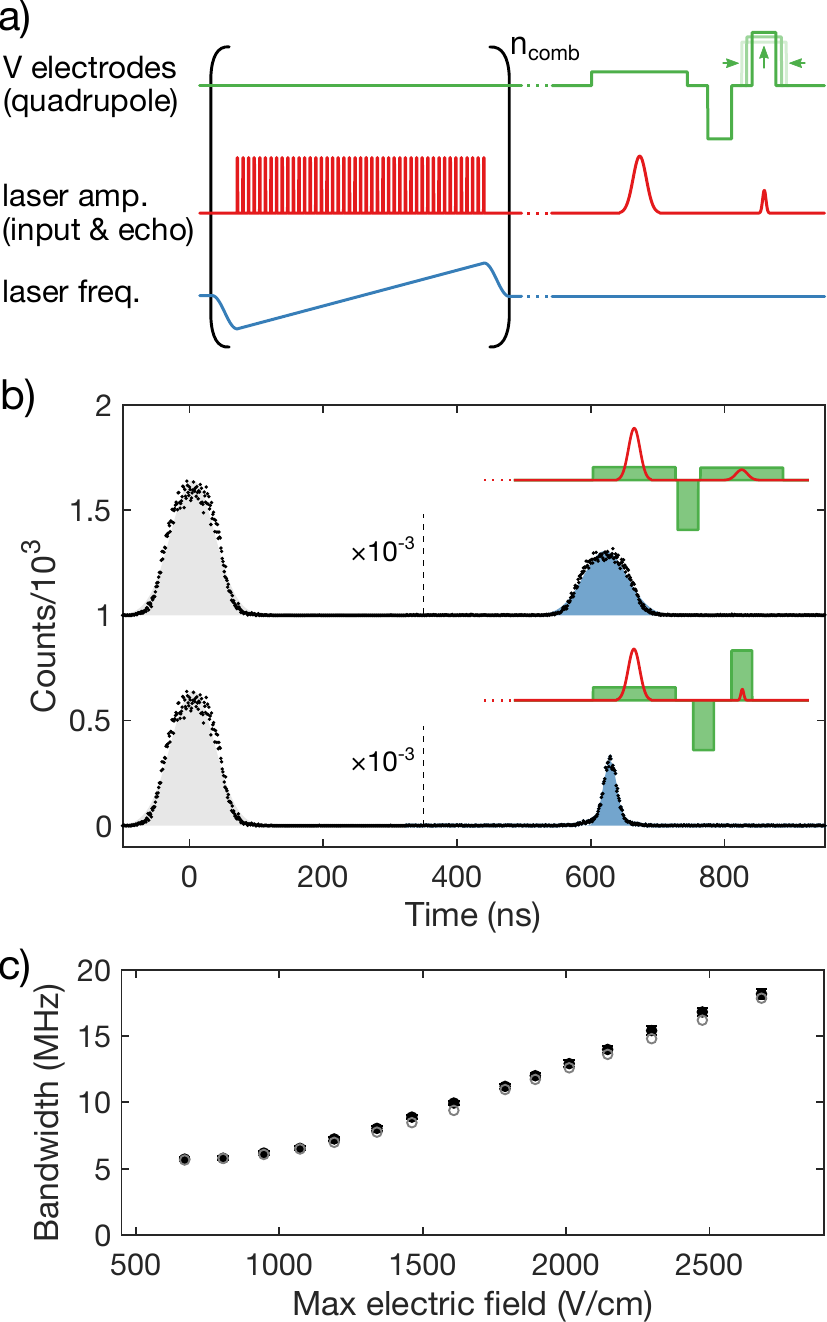}
\caption{(Color online) AFC storage with bandwidth control. a) Pulse sequence (not to scale, details in main text). b) AFC storage with (bottom) and without (top) bandwidth broadening. Colored areas are Gaussian fits to photon counts data (circles) from which widths are extracted. The partially reflected input pulse with FWHM $77.4$ ns ($5.7$ MHz FWHM in the frequency domain) is shown in grey (lighter color) in both traces at $t=0$, demagnified by a factor of $10^3$. The top trace shows the case without bandwidth broadening: \mbox{$E_\mathrm{max}(t=630\: \mathrm{ns})=E_\mathrm{max}(t=0)=0.67$ kV/cm}, where the width of the output (blue, darker) is $77.1 \pm 2.0$ ns ($5.7 \pm 0.1$ MHz). The bottom trace shows the maximum bandwidth broadening, $E_\mathrm{max}(t=630\: \mathrm{ns})=4 \times E_\mathrm{max}(t=0)=2.8$ kV/cm, where the width of the output (blue, darker) is $24.3 \pm 0.5$ ns ($18.1 \pm 0.4$ MHz). Insets show schematics of electrode pulse sequences. c) Bandwidth of pulses as a function of the $E^\mathrm{max}_\mathrm{output}$. In all cases, $E^\mathrm{max}_\mathrm{input}=0.67$ kV/cm. Filled black circles are FWHM data. Error bars, which are smaller than the markers, represent $95\%$ confidence intervals from fits. Unfilled grey circles are simulation data (see main text and Appendix \ref{sec:simulation}).}\label{fig:bw}
\end{figure}

The bandwidth of stored light can be dynamically controlled by biasing the electrodes in a quadrupole configuration as shown in the bottom panel of Figure \ref{fig:device}a. Figure \ref{fig:bw}a shows the pulse sequence used to achieve AFC storage with bandwidth control. First, an AFC with $\Delta=1.6$ MHz and bandwidth 144 MHz is created by repeatedly burning away population $n_\mathrm{comb}=20$ times. Next, an input pulse is sent into the device, leading to an output pulse at $t=\frac{1}{\Delta}=630$ ns. In the quadrupole configuration, electric pulses create a gradient electric field across the ions so that each ion experiences a different Stark shift. Electric pulses are applied during the input and output optical pulses, and also during the wait time. The first electric field pulse, applied during the absorption of the input pulse, and the third electric field pulse, applied during the emission of stored light, induce changes to the atomic frequency comb which lead to a change in the output light frequency profile. The second pulse during the wait time is used to add phase compensation, accounting for the fact that AFC storage is first-in-first-out \cite{Usmani2010,Hosseini2009}.

Figure \ref{fig:bw}b shows AFC storage with no broadening (top) and with the maximum achieved bandwidth broadening (bottom). A broadening in frequency space is seen as a narrowing of the output pulse in time. By fitting the output pulses to Gaussians, the temporal FWHMs ($\Delta t$) of input and output pulses are extracted and converted to bandwidth or frequency FWHMs ($\Delta f$) using: $\Delta f=\frac{4\mathrm{ln}2}{2\pi} (\Delta t)^{-1}$. Figure \ref{fig:bw}c shows the trend of output bandwidth as a function of the maximum electric field applied during the third pulse $E_\mathrm{max}$ (the electric field across the resonator ranges from $-E_\mathrm{max}$ to $E_\mathrm{max}$). To confirm that the trend observed in the data is expected given the atomic frequency comb profile, the input pulse, and the electric field distribution $E_y(x)$, a simulation of the experiment was performed by numerically integrating the time-evolution equations of the atoms and cavity (see Appendix \ref{sec:simulation}). The simulation data reproduces the trend in FWHM as a function of field. The only previously unknown parameter used in this simulation was the distance that the optical mode penetrates into the photonic crystal mirrors, which modifies the effective resonator length and changes the value of $E_\mathrm{max}$. This parameter was found to be $x_\mathrm{eff}=6$ $\mu$m for each mirror by coarsely sweeping $x_\mathrm{eff}$ in \mbox{1 $\mu$m} increments in the simulation to find the best fit to the data.

\section{Discussion}

In this work, we have demonstrated the capabilities of an on-chip optical storage device with DC Stark shift control. Taking this technology on-chip has two main advantages. First, it allows miniaturization and future integration with other optical components on chip. Second, it enables simple generation of large electric fields. Because the distance between electrodes across the resonator is small (a minimum of 20 $\mu$m), electric fields of $3$ kV/cm are generated with just $\pm5$ V of applied bias in the parallel configuration. Such biases were easily supplied by a function generator with no additional amplification. In the quadrupole configuration, electric field gradients of up to 50 V/cm/$\mu$m were generated, corresponding to gradient of 0.58 MHz/$\mu$m in \Er:\YSO transition frequencies.

For the dynamically controlled memory times in Fig$.$ \ref{fig:killnr}d, an excellent match was found between the amplitude of stored light as a function of time and the theoretical limit due to the dephasing of a comb with finesse $F=12.2$, indicating that the two electric field control pulses did not introduce any irreversible dephasing. This was also confirmed using a two pulse photon echo measurement, where inserting two equal and opposite electric field pulses between the first and second optical pulses was found not to decrease the optical 
coherence time $T_2$, which was measured in this device to be $108 \pm 13\:\mu$s.

Frequency control was demonstrated for up to $\pm39$ MHz. In this work, the maximum shift was set by the maximum applied electric field of $3$ kV/cm. One technical difficulty is that ions from the other subclass that are outside of the comb will act as an absorbing background when the comb is shifted in frequency and the other subclass experiences an opposite frequency shift. Assuming that the comb can be sufficiently separated in frequency from the other subclass using high electric fields, a more fundamental limit is set by the inhomogeneity of the Stark shifts, which leads to a decrease in storage efficiency with increasing frequency shift. In this device, the Stark shift inhomogeneity was dominated by an electric field distribution that was not perfectly homogeneous (see Fig$.$ \ref{fig:device}g). Even in a perfectly homogeneous field, however, some inhomogeneity in Stark shifts will exist due to crystal field variations throughout the crystal \cite{Macfarlane2007}.

The bandwidth of stored light was changed by a factor of three from 6 MHz to 18 MHz. The maximum broadening in this case was limited by the maximum electric field gradient of 50 V/cm/$\mu$m. With higher gradients, stored pulses could be broadened up to half the value of the bandwidth of the comb, and the bandwidth of combs in this material is limited to $\sim150$ MHz \cite{Craiciu2019}. Decreasing the bandwidth of a stored pulse is not possible with this procedure, because the AFC cannot be made narrower with a gradient electric field, only wider. Narrowing the AFC could be accomplished with a frequency selective shift such as the AC Stark shift. Note that while digital memory time and frequency control theoretically do not affect the storage efficiency, the bandwidth control procedure has some associated loss. This is because the AFC rephasing does not always finish within the window defined by the third electric pulse, so the edges of the emitted pulse's temporal envelope are clipped.

An on-chip resonator allows for storage efficiencies approaching unity if the impedance matching condition is met \cite{Afzelius2010}. In this device, the storage efficiency was up to $0.4\%$, depending on the finesse of the comb created, and was mainly limited by the low coupling between the ensemble of ions and the optical mode of the resonator, characterized by an ensemble cooperativity $C<1$. The storage time on an optical transition is ultimately limited by the optical coherence time $T_2$. However, in \Er:\YSOns, superhyperfine coupling to yttrium nuclear spins in the crystal prevents the creation of narrow spectral features, which means a low storage efficiency for storage times longer than $\sim 500$ ns \cite{Craiciu2019}. Superhyperfine coupling is a major limitation to high-efficiency long lived storage in \Er:\YSO when using memory protocols based on spectral tailoring such as AFC. 

For quantum repeater applications, the duration and efficiency of on-chip storage must be improved. Improvements to the intrinsic quality factor of the resonator are required to reach the impedance matching condition. Creative solutions such as using clock transitions in \Er:\YSOns \cite{Businger2020,McAuslan2012,Rakonjac2020}, which are less sensitive to superhyperfine coupling, or finding new crystal hosts for erbium ions \cite{Phenicie2019} can be used to overcome the superhyperfine limit. Another requirement of quantum memories is to store quantum states of light with high fidelity. This has already been demonstrated with the AFC protocol \cite{Saglamyurek2015}. Storage of weak coherent states using the AFC protocol with DC Stark shift control of storage time has also been recently demonstrated \cite{Horvath2020}. Future work should include demonstrations of on-chip storage of light at the quantum level with dynamic frequency and bandwidth control. More generally, this type of device could work with different absorbers that experience linear Stark shifts, or with other quantum storage protocols that do not require spectral tailoring such as Stark echo modulation memory \cite{Arcangeli2016}. 

The functionality of the device is not limited to the demonstrations in this work. For example, a gradient field could be used instead of a homogeneous field to dynamically control the storage time. The bandwidth or frequency of emissions at any time $t=\frac{m}{\Delta}$ could be modified, frequency and bandwidth control could be combined, and the order of two pulses could be reversed. A device which enables Stark shift control of an ion's transition frequency is useful for other technologies as well. For example, a gradient electric field could be used to tune two \Ers ions coupled to the same resonator into resonance with one another. This would enable entangling gates between the two ions, a key step in quantum repeater protocols using single ions \cite{Asadi2020}.

\section{Conclusion}

In this work we demonstrated a multifunctional on-chip device that can store light while dynamically modifying its storage time, frequency and bandwidth. Dynamic control of the memory time and the frequency profile of the output light was achieved via the linear DC stark shift of \Ers ions in \YSO. We demonstrated dynamic control of memory time in a digital fashion with storage times that were multiples of $50$ ns, for up to $400$ ns. The frequency of stored light was changed by up to $\pm39$ MHz, and the bandwidth of stored light was increased by up to a factor of three, from 6 MHz to 18 MHz. This on-chip platform, comprising a resonator evanescently coupled to an ensemble of atoms that experience a DC Stark shift and on-chip electrodes, can be adapted to other materials and other quantum memory protocols.

\begin{acknowledgments}
We acknowledge Phillip Jahelka for help with measuring the refractive index of amorphous silicon, Yunbin Guan for help measuring the erbium concentration, and Andrei Ruskuc and Hirsh Kamakari for building the superconducting magnets for this experiment. This work was supported by Air Force Office of Scientific Research (AFOSR) Grant
No. FA9550-18-1-0374, and the National Science Foundation (Grant No. EFRI 1741707). I.C. and J.R. acknowledge support from the Natural Sciences and Engineering
Research Council of Canada (Grants No. PGSD2-502755-
2017 and No. PGSD3-502844-2017). J.G.B. acknowledges support of the American Australian Association’s Northrop Grumman Fellowship.
\end{acknowledgments}

\appendix

\section{DC Stark Shift Measurement}\label{sec:Starkshift}

\begin{figure}
\includegraphics[width=3in]{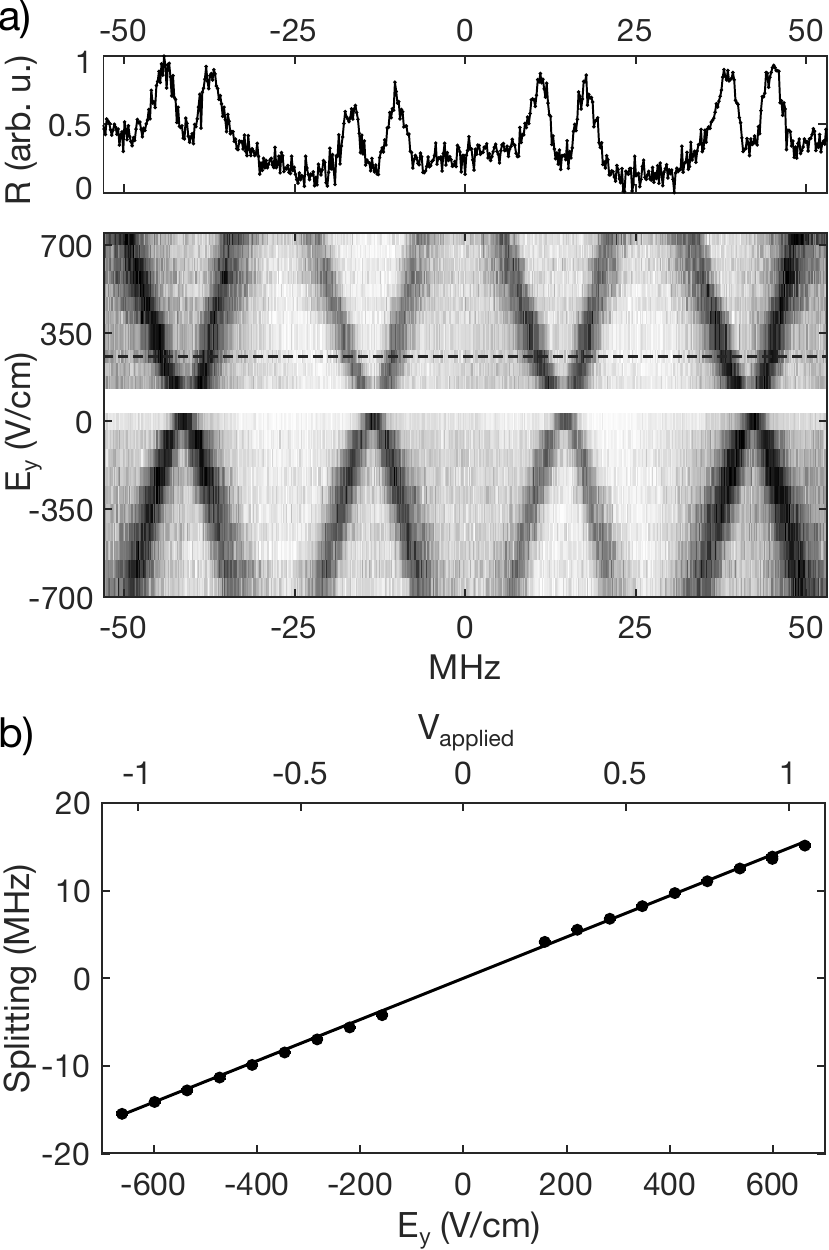}
\caption{Holeburning spectroscopy measurement of DC Stark shift. a) Frequency profile of a four-tooth comb as a function of electric field. White-black gradient represents cavity reflectance. Both subclasses of ions are present, leading to a two-fold splitting. Inset shows an example sweep with $E_y = 283$ V/cm, the location of which is indicated with a dashed line on the main plot. b) Splitting versus electric field. Black circles are splitting data obtained by fitting the traces in (a) with four sets of two-fold split Gaussians. A negative splitting means the positions of the two subclasses are reversed. Error bars, which are smaller than markers, represent $95\%$ confidence intervals from the fits. Solid line is a linear fit to the data without an offset. The top axis shows $V_\mathrm{applied}$, one output of the function generator used to generate electric pulses (the other outputs are either $V_\mathrm{applied}$ or $-V_\mathrm{applied}$). The bottom axis shows $E_y$, the electric field applied along the \YSO $b$-axis, calibrated using the average value from simulation $E_y=314.8$ V/cm when $\pm 1$ V is applied to each electrode (dashed green line in Fig$.$ \ref{fig:device}e). There is an additional factor of 2 in the calibration due to the impedance mismatch between the 50 $\Omega$ output of the function generator and the open-circuit electrodes.}\label{fig:AppendixA}
\end{figure}

The Stark shift parameter for electric fields applied along the \YSO crystal $b$-axis was estimated in the same device using spectral holeburning with electrodes biased in the parallel configuration (see Fig$.$ \ref{fig:device}e). A comb consisting of four narrow teeth 27.5 MHz apart was created in the \Er:\YSO optical transition. The frequency profile of the transition was then measured while a variable electric field was applied, which led to a field-dependent twofold splitting of each tooth, as shown in Figure \ref{fig:AppendixA}a. This splitting results from the equal and opposite Stark shifts experienced by the two subclasses of \Er:\YSO ions $\delta f_\pm = \pm s_b E$. Figure \ref{fig:AppendixA}b shows the observed splitting as a function of electric field. The slope of the straight-line fit is $2s_b$, from which $s_b=11.8 \pm 0.2$ kHz/(V/cm) is extracted. This uncertainty does not take into account any misalignment between the electric field and the $b$-axis. To align $E_y$ to the $b$-axis, the device coordinate axes ($x$ and $y$) were visually aligned to the \YSO chip edges ($b$ and $D_2$ crystal axes), with an estimated error of $<5^\circ$.

Figure \ref{fig:AppendixA}a also shows a slight broadening of the teeth as a function of electric field. This broadening is caused by the deviation of the electric field profile from a perfectly uniform field (see Fig$.$ \ref{fig:device}e) and an additional measured inhomogeneity in the Stark shifts. To quantify these effects, the expected broadening was simulated by considering the optical frequencies of a large number of ions sampled from a comb tooth, with each ion experiencing an electric field sampled from the distribution in Fig$.$ \ref{fig:device}e, and a slightly different Stark shift parameter $s_b$ sampled from a Gaussian distribution with FWHM $\gamma_s$. After accounting for the known electric field inhomogeneity, we estimated the additional inhomogeneity to be characterized by $\gamma_s=2.5$ kHz/(V/cm). The additional broadening could be the result of inhomogeneity in the Stark shift parameters \cite{Macfarlane2007} and some additional electric field inhomogeneity unaccounted for in simulation. The latter could be caused by fabrication imperfections or stray fields from other devices on the chip.

\section{Amplitude of the $\mathbf{m}$\textsuperscript{th} light emission}\label{sec:efficiency_theory}
The following section expands upon the analysis by Afzelius et al$.$ in References \cite{Afzelius2009,Afzelius2010} to consider multiple emissions. An ensemble of ions is coupled to a cavity with field decay rate $\kappa$. An AFC is created in this ensemble of ions, which leads ion distribution to be $n(\omega)$, where $\int n(\omega)
d\omega=N$, and $N$ is the number of ions. Each ion has a detuning $\omega$ relative to cavity center frequency, coherent decay rate $\gamma_{h}$, and ion-cavity coupling rate $g$. After sending a photon into the cavity that is resonant with it, the dynamic equations \cite{Afzelius2010,JB} of cavity field $\mathcal{E}$ and atomic polarization $\sigma_{\omega}$ in the rotating frame of photon frequency are 
\begin{equation} \label{Caf}
\dot{\mathcal{E}}=-\kappa \mathcal{E} + \sqrt{2\kappa_\mathrm{in}}\mathcal{E}_\mathrm{in} + ig\int n(\omega)\sigma_{\omega}d\omega,
\end{equation}
\begin{equation} \label{atomicpo}
    \dot{\sigma}_{\omega}=-(i\omega+\gamma_{h})\sigma_{\omega}+ig\mathcal{E}.
\end{equation} 
The input output formalism gives
\begin{equation} \label{inout}
    \mathcal{E}_\mathrm{out}= -\mathcal{E}_\mathrm{in}+\sqrt{2\kappa_\mathrm{in}}\mathcal{E}.
\end{equation}
where $\kappa_\mathrm{in}$ is the cavity decay rate to the input channel.
We can solve Eq. (\ref{atomicpo}) and get
\begin{equation} \label{APso}
    \sigma_{\omega}(t)=ig\int_{-\infty}^{t}e^{-(i\omega+\gamma_{h})(t-t^{\prime})}\mathcal{E}(t^{\prime})dt^{\prime}.
\end{equation}
Then, inserting Eq. (\ref{APso}) into Eq. ({\ref{Caf}}), we find
\begin{equation} \label{Cavin}
\begin{aligned}
    \dot{\mathcal{E}}(t)&= -\kappa \mathcal{E}(t) + \sqrt{2\kappa_\mathrm{in}}\mathcal{E}_\mathrm{in}(t) \\ &-g^2\int_{-\infty}^{t}e^{-\gamma_{h}(t-t^{\prime})}\int n(\delta)e^{-i\omega(t-t^{\prime})}d\omega \mathcal{E}(t^{\prime})dt^{\prime}\\
    &= -\kappa \mathcal{E}(t) + \sqrt{2\kappa_\mathrm{in}}\mathcal{E}_\mathrm{in}(t) \\
    &-g^2\int_{-\infty}^{t}e^{-\gamma_{h}(t-t^{\prime})}\widetilde{n}(t-t^{\prime}) \mathcal{E}(t^{\prime})dt^{\prime}
\end{aligned}
\end{equation}
where $\widetilde{n}(t)$ is the Fourier transform of $n(\omega)$\cite{Afzelius2009}.

We have an atomic frequency comb with period $\Delta$, and each tooth has a shape described by $f(\omega)$, so $n(\omega)$ can be written as
\begin{equation}
\begin{aligned}
    n(\omega)&\propto \sum_{k=-\infty}^{+\infty}f(\omega-k\Delta)\\
    &\propto \sum_{k=-\infty}^{+\infty}f(\omega)*\delta(\omega-k\Delta)
\end{aligned}
\end{equation}
The Fourier transform of $n(\omega)$ is $\widetilde{n}(t)$:
\begin{equation} \label{FFT}
\begin{aligned}
    \widetilde{n}(t)&\propto \sum_{k=-\infty}^{+\infty}\widetilde{f}(t)\delta\left(t-\frac{k}{\Delta}\right)\\
    &\propto
    \sum_{k=-\infty}^{+\infty}\widetilde{f}(\frac{k}{\Delta})\delta\left(t-\frac{k}{\Delta}\right)
 \end{aligned}   
\end{equation}
Inserting Eq. (\ref{FFT}) into Eq. (\ref{Cavin}), we find
\begin{equation} 
\begin{aligned}
    \dot{\mathcal{E}}(t)
    &= -\kappa \mathcal{E}(t) + \sqrt{2\kappa_\mathrm{in}}\mathcal{E}_\mathrm{in}(t)-\Gamma_\mathrm{comb}\mathcal{E}(t) \\
    &-2\Gamma_\mathrm{comb}\sum_{k=1}^{+\infty}e^{-\gamma_{h}\frac{k}{\Delta}}\widetilde{f}\left(\frac{k}{\Delta}\right)\mathcal{E}\left(t-\frac{k}{\Delta}\right)
\end{aligned}
\end{equation}
where $\Gamma_\mathrm{comb}$ is the absorption rate of atomic frequency comb\cite{Craiciu2019}, and $\Gamma_\mathrm{comb}\propto g^2$.

Consider the time after the ensemble of ions absorb the light ($t>0$). There are no input pulses after $t=0$, so $\mathcal{E}_\mathrm{in}(t>0)=0$. Applying adiabatic elimination of the cavity mode ($\dot{\mathcal{E}}(t)=0$) leads to
\begin{equation} 
    \mathcal{E}(t)=-\frac{2\Gamma_\mathrm{comb}}{\kappa+\Gamma_\mathrm{comb}}\sum_{k=1}^{+\infty}e^{-\gamma_{h}\frac{k}{\Delta}}\widetilde{f}\left(\frac{k}{\Delta}\right)\mathcal{E}\left(t-\frac{k}{\Delta}\right).
\end{equation}
The cavity field at time $t=\frac{m}{\Delta}$ goes as
\begin{equation} \label{gen}
    \mathcal{E}\left(t=\frac{m}{\Delta}\right)=-\frac{2\Gamma_\mathrm{comb}}{\kappa+\Gamma_\mathrm{comb}}\sum_{k=1}^{m}e^{-\gamma_{h}\frac{k}{\Delta}}\widetilde{f}\left(\frac{k}{\Delta}\right)\mathcal{E}\left(\frac{m-k}{\Delta}\right)
\end{equation}
From this, we can see that the amplitude of the cavity field at time $t=\frac{m}{\Delta}$ is determined by the cavity field at all earlier times $t=\frac{k}{\Delta}$, where $k=0,1,...,m-1$. We can theoretically find the amplitude of the cavity field at any time, which depends on how we modulate the cavity field at previous times.
In our case, $\gamma_{h}$ is much smaller than the teeth width \cite{Craiciu2019}, so we can ignore the term $e^{-\gamma_{h}\frac{k}{\Delta}}$. We assume each tooth has Gaussian shape, which gives
\begin{equation}
\begin{aligned}
\widetilde{f}\left(\frac{k}{\Delta}\right)
&=\mathrm{exp}\left(-\frac{1}{2}\left(\frac{ k}{\Delta}\right)^2\left(\frac{\pi\gamma}{\sqrt{2\ln{2}}}\right)^2\right) \\
&=\mathrm{exp}\left(-\frac{\pi^2}{4 \mathrm{ln} 2}\frac{k^2}{F^2}\right).
\end{aligned}
\end{equation}
where $\gamma$ is the FWHM of the Gaussian peak, and \mbox{$F=\Delta/\gamma$} as we defined in the main text, Eq. (\ref{gen}) becomes
\begin{equation} \label{e1}
    \mathcal{E}\left(t=\frac{m}{\Delta}\right)=-\frac{2\Gamma_\mathrm{comb}}{\kappa+\Gamma_\mathrm{comb}}\sum_{k=1}^{m}\mathrm{exp}\left(-\frac{\pi^2}{4 \mathrm{ln} 2}\frac{k^2}{F^2}\right)\mathcal{E}\left(\frac{m-k}{\Delta}\right).
\end{equation} 
We also know from Eq. (\ref{inout}) that the amplitude of the $k^\mathrm{th}$ emission (for $k>0$) is
\begin{equation} \label{e2}
    \mathcal{E}_\mathrm{out}\left(t=\frac{k}{\Delta}\right)=
    \sqrt{2\kappa_\mathrm{in}}\mathcal{E}\left(t=\frac{k}{\Delta}\right).
\end{equation} 
At time $t=0$ we have
\begin{equation} \label{e3}
    \mathcal{E}(t=0)=\frac{\sqrt{2\kappa_\mathrm{in}}}{\kappa+\Gamma_\mathrm{comb}}\mathcal{E}_\mathrm{in}(t=0).
\end{equation}
From the above three equations, the emission at time $t=\frac{m}{\Delta}$ has the following amplitude
\begin{equation} \label{eq:sum} 
\begin{aligned}
    \mathcal{E}_\mathrm{out}\left(t=\frac{m}{\Delta}\right)
    &=-\frac{2\Gamma_\mathrm{comb}}{\kappa+\Gamma_\mathrm{comb}}\times \\
    &\left(\sum_{k=1}^{m-1}\mathrm{exp}\left(-\frac{\pi^2}{4 \mathrm{ln} 2}\frac{k^2}{F^2}\right)\mathcal{E}_\mathrm{out}\left(\frac{m-k}{\Delta}\right) \right. \\
    & \left.+\mathrm{exp}\left(-\frac{\pi^2}{4 \mathrm{ln} 2}\frac{m^2}{F^2}\right)\frac{2\kappa_\mathrm{in}}{\kappa+\Gamma_\mathrm{comb}}\mathcal{E}_\mathrm{in}(t=0)\right).
\end{aligned}
\end{equation}
The $m^\mathrm{th}$ emission is the sum of $1^{\mathrm{st}}$ to $(m-1)^{\mathrm{th}}$ emission and the input.
In the case where we don't apply electric fields to prevent any emissions, the first and second emitted field amplitudes are
\begin{widetext}
\begin{equation}
 \mathcal{E}_\mathrm{out}\left(t=\frac{1}{\Delta}\right)=-\frac{4\kappa_\mathrm{in}\Gamma_\mathrm{comb}}{(\kappa+\Gamma_\mathrm{comb})^2}\mathrm{exp}\left(-\frac{\pi^2}{4 \mathrm{ln} 2}\frac{1}{F^2}\right)\mathcal{E}_\mathrm{in}(t=0)
\end{equation}
\begin{equation}  \label{echo2}
\begin{aligned}
    \mathcal{E}_\mathrm{out}\left(t=\frac{2}{\Delta}\right)=-\frac{4\kappa_\mathrm{in}\Gamma_\mathrm{comb}}{(\kappa+\Gamma_\mathrm{comb})^2}\mathrm{exp}\left(-\frac{\pi^2}{2 \mathrm{ln} 2}\frac{1}{F^2}\right)\mathcal{E}_\mathrm{in}(t=0) \left(\mathrm{exp}\left(-\frac{\pi^2}{2 \mathrm{ln} 2}\frac{1}{F^2}\right)-\frac{2\Gamma_\mathrm{comb}}{\kappa+\Gamma_\mathrm{comb}}\right).
    \end{aligned}
\end{equation}
\end{widetext}
As Eq. (\ref{echo2}) shows, the amplitude of the second emission is composed of two parts. The first part is from the light absorbed at $t=0$, and the second part is from the light reabsorbed at the first emission time $t=1/\Delta$. The competition between these two terms determines the amplitude and the phase of the output at $t=\frac{2}{\Delta}$. When we operate in the high finesse regime (since we always want the dephasing term $\mathrm{exp}\left(-\frac{\pi^2}{2 \mathrm{ln} 2}\frac{1}{F^2}\right)$ to be close to 1), if the amplitude of the first output is small, the amplitude of the second output will be dominated by the first term in Eq. (\ref{echo2}), so it will still have an observable amplitude. If the amplitude of the first output is high, the amplitude of the second output  will be small due to the minus sign between the two terms in Eq. (\ref{echo2}). In particular, when the impedance matching condition\cite{Afzelius2010} holds where $\frac{2\Gamma_\mathrm{comb}}{\kappa+\Gamma_\mathrm{comb}}\to1$, the second emission will be zero. This trend also holds for higher order emissions, as can be seen by extending the analysis of Eq. (\ref{eq:sum}.

In the case where we apply an electric field to kill all the lower order emissions (from 1 to $m-1$), we find the $m^\mathrm{th}$ output amplitude to be
\begin{equation}
    \mathcal{E}_\mathrm{out}\left(t=\frac{m}{\Delta}\right)=-\frac{4\kappa_\mathrm{in}\Gamma_\mathrm{comb}}{(\kappa+\Gamma_\mathrm{comb})^2}\mathrm{exp}\left(-\frac{\pi^2}{4 \mathrm{ln} 2}\frac{m^2}{F^2}\right)\mathcal{E}_\mathrm{in}(t=0).
\end{equation}
Then, we can find the efficiency of the $m^\mathrm{th}$ output pulse to be
\begin{equation}
\begin{aligned}
    \eta&=\left\lvert\frac{\mathcal{E}_\mathrm{out}(t=\frac{m}{\Delta})}{\mathcal{E}_\mathrm{in}(t=0)}\right\rvert^2\\
    &=\left(\frac{\kappa_\mathrm{in}}{\kappa}\frac{4\Gamma_\mathrm{comb}/\kappa}{(1+\Gamma_\mathrm{comb}/\kappa)^2}\right)^2\mathrm{exp}\left(-\frac{\pi^2}{2 \mathrm{ln} 2}\frac{m^2}{F^2}\right).
\end{aligned}
\end{equation}

\section{Time evolution simulations}\label{sec:simulation}

Simulations of the cavity-ion system were performed to confirm the trends observed in the bandwidth control experiment. These simulations involved numerically solving the discrete form of Eq. (\ref{Caf}) and Eq. (\ref{atomicpo}) for the cavity field $\mathcal{E}$ and the atomic polarization $\sigma_{i}$ of a number $n$ of ions in the rotating frame, following Reference \cite{Diniz2011}:
\begin{equation}\label{eq:cavity_sim}
    \dot{\mathcal{E}}(t)=-\left(\kappa+i\Delta \omega_a \right)\mathcal{E}(t) + \sqrt{2\kappa_\mathrm{in}}\mathcal{E}_\mathrm{in}(t)+i \sum_{i=1}^{n} g \sigma_i(t) ,
\end{equation}
\begin{equation}\label{eq:ion_sim}
    \dot{\sigma}_i(t)=-\left(\gamma_h+i\Delta \omega_i(t) \right)\sigma_i(t) + ig \mathcal{E}(t),
\end{equation}

where $\Delta \omega_a$ is the cavity detuning and $\Delta \omega_i(t)$ is the detuning of each ion, which can vary in time as a function of applied electric field at the location of the ion $\Delta \omega_i(t) = \Delta \omega_{i,0} \pm s E_y(x_i,t)$. $\Delta \omega_{i,0}$ is the detuning of each ion in the absence of an applied electric field and the $\pm$ sign depends on which subclass the ion is in. The cavity field is coupled to external fields as described by input-output formalism (see Equation \ref{inout}). The initial conditions are $\mathcal{E}(0)=0$, $\sigma_i(0)=0$.

For the simulation, a system of $n+1$ differential equations (Equations \ref{eq:ion_sim} and \ref{eq:cavity_sim}) are numerically solved. To keep the number of equations to a reasonable size, the number of ions simulated $n \sim 10^4$ is significantly smaller than the true number of ions coupled to the cavity $\sim 10^7$. To accurately represent the strength of the interaction between the ion ensemble and the cavity, $g$ in the simulation is chosen such that $g_\mathrm{total}^2=ng^2$, where $g_\mathrm{total}=2\pi \times 0.6$ GHz is measured from the cavity reflectance curve \cite{Craiciu2019}. The time-independent frequency distribution of the ions (frequency comb) is described as a continuous distribution, and $n$ values of $\Delta \omega_{i,0}$ are sampled from it. A time dependent scalar $\pm s E_y(x_i,t)$ representing the Stark shift is added to all ion detunings. $E_y(x_i,t)$ for each ion is given by randomly sampling the $x$-position along the resonator, and obtaining the corresponding electric field from Figure \ref{fig:device}g, then varying it in amplitude and time to represent each electric pulse. 

Using this simulation, the output pulse profile $\mathcal{E}_\mathrm{out}(t)$ can be computed given $\mathcal{E}_\mathrm{in}(t)$, the input pulse profile centered around $t=0$, and certain set of electric field control pulses.

\bibliography{PRX_Craiciu2020_Bibliography}

\end{document}